\documentclass[prl,superscriptaddress,twocolumn,showpacs,%
amsmath,amssymb]{revtex4-1}

\usepackage{slashed}
\usepackage{graphicx}
\usepackage{color}

\newcommand{\bea}{\begin{eqnarray}}
\newcommand{\eea}{\end{eqnarray}}
\newcommand{\beq}{\begin{equation}}
\newcommand{\eeq}{\end{equation}}

\newcommand{\Md}{M_{\text{eff}}}

\newcommand{\Nc}{N_{\text{c}}}
\newcommand{\Nf}{N_{\text{f}}}
\newcommand{\tr}{\text{tr}}
\newcommand{\Tr}{\text{Tr}}
\newcommand{\diag}{\text{diag}}
\newcommand{\Det}{\text{Det}}
\newcommand{\bp}{\boldsymbol{p}}

\begin{document}
\title{Chiral gap effect in curved space}

\author{Antonino Flachi}
\affiliation{Centro Multidisciplinar de Astrof\'{i}sica,
  Departamento de F\'{i}sica, Instituto Superior T\'{e}cnico,
  Universidade de Lisboa, Avenida Rovisco Pais 1,
  1049-001 Lisboa, Portugal}
\author{Kenji Fukushima}
\affiliation{Department of Physics, The University of Tokyo,
  7-3-1 Hongo, Bunkyo-ku, Tokyo 113-0033, Japan}

\begin{abstract}
 We discuss a new type of QCD phenomenon induced in curved space.  In
 the QCD vacuum, a mass-gap of Dirac fermions is attributed to the
 spontaneous breaking of chiral symmetry.  If the curvature is
 positive large, the chiral condensate melts but a chiral invariant
 mass-gap can still remain, which we name the chiral gap effect in
 curved space.  This leads to decoupling of quark deconfinement which
 implies a view of black holes surrounded by a first-order QCD
 phase transition.
\end{abstract}
\maketitle

%%%%%%%%%%   Introduction   %%%%%%%%%%
\paragraph{Introduction.}

Quantum field theory in curved space-time is a well-established
subject~\cite{parker} with phenomenological applications to nuclear
physics and condensed matter physics now being developed.  In the
presence of a gravitational background, the metric tensor is naturally
distorted from flat Minkowskian space-time.  In usual laboratory
environments, gravitational effects are negligibly small as compared to
typical scales in the considered theory: for example,
$\Lambda_{\text{QCD}}$ in quantum chromodynamics (QCD).  (See
Refs.~\cite{Schutzhold:2002pr} for a discussion of the possibility that
such small effects could explain the QCD origin of dark energy.)

In the Universe, in contrast to laboratory experiments, it may be
possible to imagine a situation where curvature can be as large as
$\Lambda_{\text{QCD}}^2$, e.g., black holes \footnote{The scalar curvature of the Schwarzschild black hole is vanishing, but
as we explain later, we need to perform a conformal transformation to
introduce a temperature independently, and then a finite curvature arises.}. While it is often said
that ordinary quantum field theory becomes problematic in the vicinity
of the event horizon of a black hole {(i.e., the trans-Planckian
  problem)}, one can safely consider a region at a specifically chosen
distance from the horizon where {the standard model should be
  still valid under non-negligible gravitational effects.}

Because of the {complicated} vacuum structure of QCD, in order
to accommodate nonperturbative phenomena like dynamical generation of
mass and confinement of quarks and gluons, it is conceivable to regard
the black hole as an extended object surrounded by a ``media''
consisting of QCD vacuum.  Such a crustlike content around the black
hole is naturally associated with QCD phase transitions (as discussed
in Refs.~\cite{Flachi:2011sx}) and should deserve a more attentive
study along similar lines to those related to the recent disputes on
the black hole complementarity, namely, the so-called firewall
hypothesis~\cite{Almheiri:2012rt}.  At classical level, one may
think that no drastic phenomenon should happen in the freely falling
frame in which the gravitational field, even if very strong, is
locally canceled.  In QCD, however, hadron wave functions at rest and
those in the infinite momentum frame look totally
different~\cite{Susskind:1993aa}.  In fact, as we discuss later, the
QCD vacuum structure filled with quantum fluctuations should change
drastically near, but not too close to, the horizon of the black
hole.

Also in laboratory experiments, interestingly enough, descriptions
based on quantum field theory in curved backgrounds are, in some
cases, important.  In relativistic heavy-ion collisions, hot and dense
{QCD} matter (the quark-gluon plasma) is created, and it goes
though expansion at the speed of light.  Therefore, the space-time
evolution has an event horizon~\cite{Kharzeev:2005iz}, and quantum
spectra in the expanding geometry (i.e., in the Bjorken coordinates)
look analogous to those in Rindler coordinates.  {[See
  similarity between Eq.~(27) of Ref.~\cite{Dusling:2012ig} and
  Eq.~(5.25) of Ref.~\cite{'tHooft:1996tq}.]}
In this context, the speculative scenario that particle production in
QCD may be related to a Hawking temperature characterized by the
saturation scale of the strong interaction is certainly
suggestive~\cite{Castorina:2007eb}.

Effects of curved spacetime are also relevant for condensed matter
laboratory experiments. There are, for instance, theoretical proposals
(awaiting experimental confirmation) of Hawking radiation from
acoustic ``black holes'' in atomic Bose-Einstein
condensates~\cite{Garay} with the density correlation being an
experimental signature~\cite{Balbinot}.  The formulation and physical
implications of the present work may have potential relevance for such
systems of ultracold atoms.

In our study, we will focus on the effect of curvature on massless
Dirac fermions.  This problem should attract general interest not only
in relation to the chiral physics of QCD, but also for condensed
matter systems.  Massless Dirac particles are nowadays known to emerge
not only in high-energy physics: in graphene, at the interface on the
topological insulators, etc.  In principle, deformed materials may
realize a nonzero curvature in a controllable way.  Then, unlike the
case of black holes, it could be sensible to consider a negative
curvature as well (with a saddle point shaped deformation), which we
will argue later.

The following discussions are based on the observation that Dirac
fermions can have a chiral invariant mass gap due to the curvature
(and we call this the ``chiral gap effect'').  On the algebraic level
the chiral gap effect has been partially recognized for many
years~\cite{parker}, but its application to QCD is not
yet mature.  Let us look quickly over the calculation scheme to
demonstrate how such a chiral invariant mass arises.
\vspace{0.5em}

%%%%%%%%%%   Grand potential in curved space   %%%%%%%%%
\paragraph{Grand potential in curved space.}

In fermionic systems, the effective mass $\Md$ with interaction clouds
can differ from the bare one and when the Lagrangian has no explicit
mass term (i.e., chiral limit), $\Md$ should be proportional to the
scalar (chiral) condensate of the fermion bilinear:
\begin{equation}
 \Md = G \langle\bar{\psi}\psi\rangle \;.
\label{eq:mass}
\end{equation}
Here, $G$ is a coupling that adjusts the proper mass dimension.  Such
an effective mass should solve the gap equation or, equivalently,
should minimize the grand potential $\Omega[\Md]$.  Generally speaking,
the grand potential consists of two contributions: one from the tree
diagrams and the other from the fermion loops.  Namely,
$\Omega[\Md]=\Omega_{\text{tree}}[\Md]+\Omega_{\text{loop}}[\Md]$ with
\begin{equation}
 \beta\Omega_{\text{loop}}[\Md] = -\nu \ln\Det(i\slashed{\nabla}
  - \Md) \;,
\label{eq:det}
\end{equation}
where $\beta$ is the inverse temperature and $\nu$ represents the
number of fermionic degrees of freedom.  In the chiral limit, thus,
$\Md$ is the order parameter for the spontaneous breaking of chiral
symmetry.

We adopt the well-known technique of iterating the Dirac operator in
Eq.~\eqref{eq:det} in order to 
deal with a second order operator~\cite{parker}.
Then, 
we find the following:
\begin{equation}
 \beta\Omega_{\text{loop}}[\Md] = -\frac{\nu}{2}\ln\Det\biggl[\square
  + \Md^2 + \frac{R}{4} \biggr] \;.
\end{equation}
Here, $R$ is the scalar curvature.  We note that the d'Alembertian
$\square$ incorporates the spin connection.
For maximally symmetric geometries such as (anti-) de~Sitter space, it is
often possible to take the above determinant exactly (see, {for instance},
Ref.~\cite{Inagaki:1997kz} and references therein), but the final
expression is too much involved to guide plain intuition.

To extract the essence of underlying physics, it is convenient to
introduce a truncation scheme.  Let us first take an ultrastatic
metric; i.e., $g_{\tau\tau}=1$ (in the Euclidean convention).  We can
actually reach this special form by means of an appropriate conformal
transformation.  We can then adopt the resummed heat-kernel expansion
according to the Jack-Toms-Parker ansatz~\cite{Parker:1984dj}.  In
fact, the determinant has ultraviolet singularities and the
$\zeta$-function regularization, as first utilized by
Hawking~\cite{hawking}, Dowker and Critchley~\cite{dowker}, is one of
the  methods most frequently used, particularly in curved space-time {(see, however, Ref.~\cite{Ebert:2008tp} for an example that uses cutoff regularization}).
We can thus put the quantity inside the determinant into the
exponential as
\begin{equation}
 \begin{split}
 &\Tr_{\text{space}} \, e^{-t[-\partial_\tau^2 - \Delta + \Md^2 + R/4]}\\
 &\qquad = \frac{1}{(4\pi t)^2}\, e^{-t[-\partial_\tau^2
  + \Md^2 + R/4 - R/6]} \sum_k \mbox{tr}\,a_k\, t^k \;.
 \end{split}
\label{eq:heat}
\end{equation}
Here, $\Tr_{\text{space}}$ means that we take the trace over spatial
coordinates only, and $a_k$'s represent the resummed heat-kernel
coefficients as listed in Ref.~\cite{Parker:1984dj}. 
The crucial point is that Eq.~\eqref{eq:heat} makes a full resummation with respect to $R$: the coefficients $a_k$'s are functions of $R_{\mu\nu\rho\sigma}$ and $R_{\mu\nu}$ but not of $R$. The above expression Eq.~\eqref{eq:heat} is valid for both constant and nonconstant curvature spacetimes. Notice that in the presence of a cosmological constant, one is led to the former case, and our considerations with constant curvature will be valid.
The first few coefficients, for the case of constant curvature, read:
\begin{equation}
 \begin{split}
 & a_0 = 1\;, \qquad a_1 = 0\;,\\
 & a_2 = \frac{1}{180} \biggl[ R_{\mu\nu\rho\sigma}R^{\mu\nu\rho\sigma}
  - R_{\mu\nu}R^{\mu\nu} \biggr] +{1\over 12} W_{\mu\nu}W^{\mu\nu}\;,
 \end{split}
\label{eq:as}
\end{equation}
where $R_{\mu\nu\rho\sigma}$ and $R_{\mu\nu}$ are the Riemann and the
Ricci curvature tensors, respectively, and
$W_{\mu\nu} = \left[\nabla_\mu, \nabla_\nu \right]$.  
Notice that while the property of resummation is maintained also for spacetimes with nonconstant
%{inhomogeneous}
curvature, 
%but 
in this more general case, 
%while the {$R$-resummation is valid}, 
additional terms (dependent on the
derivatives of {$R$ and $W_{\mu\nu}$}) appear in $a_2$.  Such
modifications will not change the main results discussed below.  For
more calculation details, see Refs.~\cite{Flachi:2010yz}.  {We
  note that we do not take the trace over the $\gamma$ matrices in
  Eq.~\eqref{eq:heat}.  This means that $a_k$'s in Eq.~\eqref{eq:as}
  should be interpreted as matrices with respect to the Dirac
  indices.}

For the time being, we neglect $a_k$ with $k>0$ to capture the
qualitative features of fermions in curved spacetime and come back to
these corrections later.  We note that such a truncation is
justifiable if the number of dimensions, $D$, is large enough.
Although we do not assume any specific form of the geometry
except that the scalar curvature is constant, let us take simple
concrete examples.
For maximally symmetric geometries such as (anti-) de~Sitter space, indeed,
the Riemann and the Ricci tensors are as suppressed by $D$ as
\begin{equation}
 \frac{R_{\mu\nu\rho\sigma}R^{\mu\nu\rho\sigma}}{R^2}
  = \frac{2}{D(D-1)}\;, \qquad
 \frac{R_{\mu\nu}R^{\mu\nu}}{R^2} = \frac{1}{D} \;.
\label{eq:suppressD}
\end{equation}
For more general geometries, one may come to the same conclusion by
using the Weyl decomposition (see formula (228) in
Ref.~\cite{chandra}).

Once we admit the $k=0$ dominance, we eventually come by a very simple
picture, where the effect of the scalar curvature replaces the
effective mass as seen in Eq.~\eqref{eq:heat} as
\begin{equation}
 \Md^2 \;\to\; \Md^2 + \frac{R}{12} \;.
\label{eq:mass_shift}
\end{equation}
This is a profound observation, and at the same time, seems to be
puzzling at a glance.  In the chiral symmetric phase, we have $\Md=0$,
but fermions are still gapped with $R/12$.  How can we reconcile
chiral symmetry and such gapped fermions?
\vspace{0.5em}

%%%%%%%%%%   Similarity and dissimilarity to the thermal mass   %%%%%%%%%
\paragraph{Similarity and dissimilarity to the thermal mass.}

This kind of chiral invariant mass is common in thermal field theory.
In the hard thermal loop resummation, the self-energy should be
inserted in the fermion propagator which produces a thermal
mass~\cite{lebellac}; nevertheless, it does not affect the
commutativity between the propagator and $\gamma_5$.  In the limit of
vanishing spatial momenta ($\bp=0$), the fermion propagator inverse
reads:
\begin{equation}
 i S^{-1}(p_0) =
  p_0 \gamma^0 - \frac{m_T^2}{p_0}\gamma^0 - \Md \;,
\end{equation}
as a function of the energy $p_0$, where $m_T^2=(1/6)g^2 T^2$
represents the one-loop thermal mass squared.  The pole is located at
$p_0=\pm m_T\neq 0$, even when $\Md=0$ in the chiral symmetric
phase.  Although $m_T^2$ must be a function of $T$, the authors of
Ref.~\cite{Hidaka:2006gd} have calculated $\Omega[\Md]$ as a function
of $m_T$ as if $m_T$ is an independent variable and they found the
chiral transition temperature lowered with increasing
$m_T$.

We see that the correction to the fermion mass due to $R$ is quite
similar to $m_T^2$.  As a matter of fact, the interpretation of
Eq.~\eqref{eq:mass_shift} is even simpler.  If we na\"{i}vely combine
it with Eq.~\eqref{eq:mass}, a shift in
$G^2\langle\bar{\psi}\psi\rangle^2$ would break chiral symmetry.  We
should, however, identify this shift in a chiral invariant $G^2\rho$
with $\rho\equiv\langle\bar{\psi}\psi\rangle^2
 + \langle\bar{\psi}i\gamma_5\boldsymbol{\tau}\psi\rangle^2$.  Hence,
the fermion mass gap can be consistent with chiral symmetry.

This mechanism to generate a mass-gap helps us to understand how the
curvature would affect the chiral phase transition. 
Let us consider a spacetime of constant curvature and parametrize the
$\rho$-dependence of the grand potential as
%The grand potential should be a function of $\rho$ and we can parametrize it as
\begin{equation}
 \beta\Omega[\rho] = a(T-T_c)\rho + \lambda\rho^2 + \cdots
\end{equation}
near the second-order critical point at $T=T_c$. {It should
  be noted that we can introduce the temperature $T$ independently from
  $R$ by using the metric tensors depending on the real-time $t$
  and compactifying the manifold along the imaginary-time $\tau$.}
In the presence of a finite curvature, Eq.~\eqref{eq:mass_shift}
shifts $\rho$ as $\rho+R/(12G^2)$, and so $T_c$ is also modified as
\begin{equation}
 \beta\Omega[\rho] \;\to\; \biggl[ a(T-T_c) + \frac{\lambda R}{6G^2}
  \biggr]\rho + \rho^2 + \cdots \;,
\end{equation}
from which the critical temperature shifts to
\begin{equation}
 T_c^\ast = T_c - \frac{\lambda R}{6G^2 a} \;.
\label{eq:tc}
\end{equation}
This is a result reminiscent of the effect of $m_T$ on the chiral
phase transition as argued in Ref.~\cite{Hidaka:2006gd}.  We note that
this shift successfully reproduces the qualitative behavior found in
solvable examples~\cite{Inagaki:1997kz}.

We shall point out two important differences between the roles played
by $m_T$ and $R$.  The first is that, unlike $m_T$, $R$ is an
independent variable and is not constrained by $T$.  Curvature and
temperature effects are in sharp contrast {in the way they manifest}
%for \commt{the manifestation} 
in realistic systems: we cannot address a quantum
phase transition induced by $m_T$, but it would be sensible to do so
for $R$.  In fact, Eq.~\eqref{eq:tc} implies that a phase transition
takes place at $R=6G^2 a T_c/\lambda$ even for $T=0$.

The second difference is that a negative shift with $R<0$ is also
possible in Eq.~\eqref{eq:mass_shift} (see Ref.~\cite{gorbar} for 
an analysis of dynamical symmetry breaking in spaces of negative curvature).  
Then the effective mass increases by $|R|/12$, so that $T_c^\ast$ should increase.  This may
lead us to an interesting conjecture that the chiral phase transition
and the quark deconfinement could become completely distinct if $R$ is
negative large.  To confirm this, however, we need to address the
Yang-Mills dynamics in curved spacetime using a first-principle
approach~\cite{Yamamoto:2014vda}, which is beyond our current scope.
Instead, in this work, we shall focus on the effect of fermion
excitations on the quark deconfinement {and find that the
  decoupling tendency is common also for $R>0$.}
\vspace{0.5em}

%%%%%%%%%%   Thermal excitation and quark deconfinement   %%%%%%%%%
\paragraph{Thermal excitation and quark deconfinement.}

At finite $T$, we can characterize the quark deconfinement using an
order parameter:
\begin{equation}
 \Phi = \frac{1}{\Nc}\tr L\;,
\end{equation}
which is called the (traced) Polyakov loop (see
Refs.~\cite{Fukushima:2011jc} and references therein).  We can
rigorously define the quark deconfinement only in the pure Yang-Mills
theory that has center symmetry.  In QCD with light flavors the
deconfinement crossover turns out to be smooth due to fermion
interactions.  As discussed in Ref.~\cite{Fukushima:2003fw}, the
chiral phase transition controls the fermion mass, and the
deconfinement would be more favored with lighter quarks after the
chiral phase transition.  This is a qualitative picture to understand
the simultaneous crossover of quark deconfinement and chiral
restoration as observed in the lattice QCD simulation.

Thermally excited fermions on the gluonic background generate terms
that break center symmetry.  They concretely arise from
\begin{equation}
 \beta\Omega_{\text{loop}}[\Md] = - \Nf\sum_{i=1}^{\Nc} \ln\Det
  \bigl( i\slashed{\nabla} - \Md + i\phi_i \gamma^t \bigr)\;,
\label{eq:quark_poten}
\end{equation}
where $\nu$ is specified as the flavor number $\Nf$, and we also take
the trace over color up to $\Nc$.  In the determinant a new variable
$\phi_i$'s appear to represent the eigenvalues of the Polyakov loop
matrix: $L=\diag(e^{i \phi_i}) \quad (i=1,2,\dots,\Nc)$.  In a special
gauge called the Polyakov gauge, $\phi_i$'s correspond to the diagonal
components of the temporal gauge potential $A_\tau$.

As we already saw, the effective mass is shifted by $R/12$, and a
straightforward summation over the Matsu\-bara frequencies yields
$\Omega_{\text{loop}}^{T=0}+\Omega_{\text{loop}}^T$ with
\begin{equation}
 \begin{split}
 & \beta\Omega_{\text{loop}}^T = -2\Nf\, V \int\frac{d^3 p}{(2\pi)^3}
  \Tr \Bigl[ \ln\bigl( 1+L\,e^{-\beta(\varepsilon_p - \mu)} \bigr) \\
  &\qquad\qquad\qquad\qquad\qquad
   + \ln\bigl( 1+L^\dag\,e^{-\beta(\varepsilon_p + \mu)} \bigr)
   \Bigr]\;,
 \end{split}
\end{equation}
where the quasiparticle energy dispersion is
$\varepsilon_p\equiv\sqrt{p^2+\Md^2+R/12}$.  This is a simple
expression but it encompasses the essence of all complicated
calculations as done in Refs.~\cite{Sasagawa:2012mn,Flachi:2013zaa}.
In flat space, usually, $\Md$ controls the explicit breaking of center
symmetry.  As soon as a nonzero $R$ is turned on, thermally excited
fermions are suppressed by not only $\Md$ but also $R$.  Therefore,
even in the chiral symmetric limit, if $R$ is larger than $T$, fermion
excitations are almost absent, so that center symmetry can be an
approximate symmetry.

%---   figure   ---%
\begin{figure}
 \includegraphics[width=\columnwidth]{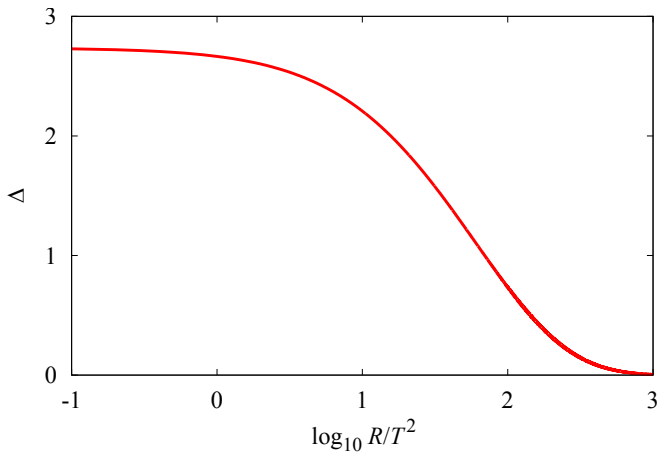}
 \caption{{Coupling strength between dynamical quarks and the
     gluonic sector quantified through the center symmetry breaking
     $\Delta$ as a function of dimensionless curvature $R/T^2$.}}
 \label{fig:pol}
\end{figure}
%---   figure   ---%

To quantify this speculation, let us plot the (dimensionless)
magnitude of the center symmetry breaking for $\mu=0$; i.e.,
$\Delta\equiv (\beta^4/V)\bigl(\Omega_{\text{loop}}^T[\Phi=1]
-\Omega_{\text{loop}}^T[\Phi=-1]\bigr)$.  As we see in
Fig.~\ref{fig:pol}, because of a small coefficient $1/12$, we need to
have hundreds times as large $R$ as $T^2$ to realize decoupling.

Once the decoupling happens, the gluonic sector should behave as
pure Yang-Mills theory.  Thus, the quark deconfinement transition
should be of first order rather than smooth crossover.
{One may think that the deconfinement is also eased by large
  $R$, and indeed, we remark that the infrared singularity is weakened
  in curved spacetime~\cite{Elizalde:1996am}.}
\vspace{0.5em}

%%%%%%%%%%   Higher-order corrections   %%%%%%%%%
\paragraph{Higher-order corrections.}

Higher-order corrections {from the heat kernel expansion} can be
easily taken into account in our scheme.  These terms involve
combinations of Riemann and Ricci curvature tensors and correct the
grand potential by $\delta\Omega_{\text{loop}}$.  We can utilize
Eq.~\eqref{eq:as} to show that it is related to the leading-order term
as
\begin{equation}
 \delta\Omega_{\text{loop}} = 
%-\frac{R_{\mu\nu\rho\tau}^2}{96}
  a_2  \biggl(\frac{\partial}{\partial\Md^2}\biggr)^2
  \Omega_{\text{loop}} \;.
\label{eq:Omega_shift}
\end{equation}
It is a nontrivial finding that the correction terms take a form of
mass derivatives.    Then, we notice that a mass shift can reproduce
the above result as
$\Omega_{\text{loop}}[\Md^2+\delta\Md^2] \approx
 \Omega_{\text{loop}}[\Md^2] + \delta\Omega_{\text{loop}}[\Md^2]$.
 {We can show that the coefficient $a_2$ is negative in general,
   and then the mass squared correction turns out to be purely
   imaginary:}
\begin{equation}
 \delta\Md^2 = i \sqrt{\left|a_2\right|} \;,
\end{equation}
or the self-energy has an imaginary part.  {Actually, one can
  confirm $a_2<0$ by plugging Eq.~\eqref{eq:suppressD} into $a_2$ in
  Eq.~\eqref{eq:as}.}
The appearance of complex energy dispersion indicates that the vacuum
is not stable.  In fact, {the curvature induced particle
  production, as observed in Ref.~\cite{Yamamoto:2014vda}, suggests an
  alteration of the vacuum persistence.}  Further investigations to
deepen our understanding on the physical interpretation of these
complex corrections is necessary.

Finally, let us also point out that Eq.~\eqref{eq:Omega_shift} is
proportional to the chiral susceptibility $\chi$.  Since $\chi$ is
enhanced at the chiral phase transition, there may be an interesting
interplay between the chiral dynamics and the curvature effect at
critical point.
\vspace{0.5em}

%%%%%%%%%%   Discussions and summary   %%%%%%%%%%
\paragraph{Discussions and summary.}

Our analysis indicates that the predominant effect on fermions in
curved space is the appearance of a chiral symmetric mass-gap due to
the scalar curvature $R$, which we call the chiral gap effect.  We
have shown that the mass shift is systematically formulated in a form
of the resummed expansion with respect to the Riemann and the Ricci
curvature tensors.  The chiral gap effect gives an intuitive
explanation for the nature of the chiral phase transition in curved
space;  chiral symmetry tends to get restored with $R>0$, while the
chiral condensate and the chiral transition temperature becomes larger
with $R<0$.  Importantly, the chiral gap effect also suggests
decoupling between the chiral dynamics and the quark deconfinement.

In principle, lattice QCD simulations can verify our conjecture.
So far, however, it is not easy to formulate the problem numerically
for a geometry that constantly curves in space.  The difficulty
originates from the singularity associated with polar coordinates
that are most convenient to describe curved geometries.  In this sense,
therefore, our analysis may be useful to guide future attempts to
simulate QCD in curved space
{or specifically in the Schwarzchild metric}.

In the future, it will be indispensable to study the gluonic sector
{more carefully} in curved space, for which lattice simulations
are the most powerful tool, but not necessarily a unique choice.  One
might utilize the strong-coupling expansion or employ a description based on the
inverted Weiss potential~\cite{Braun:2007bx} with nonflat metric
tensors, which would be an intriguing research subject to pursue.

\acknowledgments

We thank Arata Yamamoto (K.F.) and V. Vitagliano and A. Nerozzi (A.F.)
for discussions.  This work was partially supported by JSPS KAKENHI
Grant No. 24740169 (K.F.) and by the the Funda\c{c}\~{a}o para a
Ci\^{e}ncia e a Tecnologia of Portugal (FCT) and the Marie Curie
Action COFUND of the European Union Seventh Framework Program Grant
Agreement No. PCOFUND-GA-2009-246542 (A.F.).


\begin{thebibliography}{1}%
\makeatletter
\providecommand \@ifxundefined [1]{%
 \@ifx{#1\undefined}
}%
\providecommand \@ifnum [1]{%
 \ifnum #1\expandafter \@firstoftwo
 \else \expandafter \@secondoftwo
 \fi
}%
\providecommand \@ifx [1]{%
 \ifx #1\expandafter \@firstoftwo
 \else \expandafter \@secondoftwo
 \fi
}%
\providecommand \natexlab [1]{#1}%
\providecommand \enquote  [1]{``#1''}%
\providecommand \bibnamefont  [1]{#1}%
\providecommand \bibfnamefont [1]{#1}%
\providecommand \citenamefont [1]{#1}%
\providecommand \href@noop [0]{\@secondoftwo}%
\providecommand \href [0]{\begingroup \@sanitize@url \@href}%
\providecommand \@href[1]{\@@startlink{#1}\@@href}%
\providecommand \@@href[1]{\endgroup#1\@@endlink}%
\providecommand \@sanitize@url [0]{\catcode `\\12\catcode `\$12\catcode
  `\&12\catcode `\#12\catcode `\^12\catcode `\_12\catcode `\%12\relax}%
\providecommand \@@startlink[1]{}%
\providecommand \@@endlink[0]{}%
\providecommand \url  [0]{\begingroup\@sanitize@url \@url }%
\providecommand \@url [1]{\endgroup\@href {#1}{\urlprefix }}%
\providecommand \urlprefix  [0]{URL }%
\providecommand \Eprint [0]{\href }%
\providecommand \doibase [0]{http://dx.doi.org/}%
\providecommand \selectlanguage [0]{\@gobble}%
\providecommand \bibinfo  [0]{\@secondoftwo}%
\providecommand \bibfield  [0]{\@secondoftwo}%
\providecommand \translation [1]{[#1]}%
\providecommand \BibitemOpen [0]{}%
\providecommand \bibitemStop [0]{}%
\providecommand \bibitemNoStop [0]{.\EOS\space}%
\providecommand \EOS [0]{\spacefactor3000\relax}%
\providecommand \BibitemShut  [1]{\csname bibitem#1\endcsname}%
\let\auto@bib@innerbib\@empty
%</preamble>
\bibitem [{Note1()}]{Note1}%
  \BibitemOpen
  \bibinfo {note} {The scalar curvature of the Schwarzschild black hole is
  vanishing, but as we explain later, we need to perform a conformal
  transformation to introduce a temperature independently, and then a finite
  curvature arises.}\BibitemShut {Stop}%
\end{thebibliography}%


\begin{thebibliography}{99}
\bibitem{parker}
 L.~Parker and D.~Toms,
 ``\textit{Quantum Field Theory in Curved Spacetime:
 Quantized Fields and Gravity},'' (Cambridge University Press, Cambridge, England, 2009);
 I.L. Buchbinder, S.D. Odintsov, and I.L. Shapiro,
 ``\textit{Effective Action in Quantum Gravity},'' (IOP press, Bristol, 1992).


\bibitem{Schutzhold:2002pr} 
  R.~Schutzhold,
  %``Small cosmological constant from the QCD trace anomaly?,''
  Phys.\ Rev.\ Lett.\  {\bf 89}, 081302 (2002);
%  [gr-qc/0204018].
%\bibitem{Urban:2009yg} 
  F.~R.~Urban and A.~R.~Zhitnitsky,
  %``The QCD nature of Dark Energy,''
  Nucl.\ Phys.\  {\bf B835}, 135 (2010).
%  [arXiv:0909.2684 [astro-ph.CO]].

\bibitem{Flachi:2011sx} 
  A.~Flachi and T.~Tanaka,
  %``Chiral Phase Transitions around Black Holes,''
  Phys.\ Rev.\ D {\bf 84}, 061503 (2011);
%  [arXiv:1106.3991 [hep-th]].
%\bibitem{Flachi:2013iia} 
  A.~Flachi,
  %``Deconfinement transition and Black Holes,''
  Phys.\ Rev.\ D {\bf 88}, 041501 (2013).
%  [arXiv:1305.5348 [hep-th]].

\bibitem{Almheiri:2012rt} 
  A.~Almheiri, D.~Marolf, J.~Polchinski and J.~Sully,
  %``Black Holes: Complementarity or Firewalls?''
  J. High Energy Phys. 02 (2013) 062;
%  [arXiv:1207.3123 [hep-th]].
  c.f. S. L. Braunstein, S. Pirandola, and K. Zyczkowski, 
  %"Better Late than Never: Information Retrieval from Black Holes," 
  Phys. Rev. Lett. {\bf 110}, 101301 (2013).


\bibitem{Susskind:1993aa} 
  L.~Susskind,
  %``Strings, black holes and Lorentz contraction,''
  Phys.\ Rev.\ D {\bf 49}, 6606 (1994).
%  [hep-th/9308139].

\bibitem{Kharzeev:2005iz} 
  D.~Kharzeev and K.~Tuchin,
  %``From color glass condensate to quark gluon plasma through the event horizon,''
  Nucl.\ Phys.\  {\bf A753}, 316 (2005)
%  [hep-ph/0501234].

\bibitem{Castorina:2007eb} 
  P.~Castorina, D.~Kharzeev, and H.~Satz,
  %``Thermal Hadronization and Hawking-Unruh Radiation in QCD,''
  Eur.\ Phys.\ J.\ C {\bf 52}, 187 (2007).
%  [arXiv:0704.1426 [hep-ph]].

\bibitem{Dusling:2012ig} 
  K.~Dusling, T.~Epelbaum, F.~Gelis, and R.~Venugopalan,
  %``Instability induced pressure isotropization in a longitudinally expanding system,''
  Phys.\ Rev.\ D {\bf 86}, 085040 (2012).
%  [arXiv:1206.3336 [hep-ph]].

\bibitem{'tHooft:1996tq} 
  G.~'t Hooft,
  %``The Scattering matrix approach for the quantum black hole: An Overview,''
  Int.\ J.\ Mod.\ Phys.\ A {\bf 11}, 4623 (1996).
%  [gr-qc/9607022].

\bibitem{Garay}
  L.J.~Garay, J.R.~Anglin, J.I.~Cirac, and P.~Zoller,
  %``Sonic black holes in dilute Bose-Einstein condensates,''
  Phys.\ Rev.\ A {\bf 63}, 023611 (2001).

\bibitem{Balbinot}
  R.~Balbinot, A.~Fabbri, S.~Fagnocchi, A.~Recati, and I.~Carusotto,
  %``Nonlocal density correlations as a signature of Hawking radiation
  %from acoustic black holes,''
  Phys.\ Rev.\ A {\bf 78}, 021603 (2008).

\bibitem{Inagaki:1997kz} 
  T.~Inagaki, T.~Muta, and S.~D.~Odintsov,
  %``Dynamical symmetry breaking in curved space-time: Four fermion interactions,''
  Prog.\ Theor.\ Phys.\ Suppl.\  {\bf 127}, 93 (1997).
%  [hep-th/9711084].


\bibitem{Parker:1984dj} 
  L.~Parker and D.~J.~Toms,
  %``New Form for the Coincidence Limit of the Feynman Propagator, or Heat Kernel, in Curved Space-time,''
  Phys.\ Rev.\ D {\bf 31}, 953 (1985);
%\bibitem{Jack:1985mw} 
  I.~Jack and L.~Parker,
  %``Proof of Summed Form of Proper Time Expansion for Propagator in Curved Space-time,''
  Phys.\ Rev.\ D {\bf 31}, 2439 (1985).

\bibitem{hawking} 
  S.~W.~Hawking,
  %``Zeta Function Regularization of Path Integrals in Curved Space-Time,''
  Commun.\ Math.\ Phys.\ {\bf 55}, 133 (1977).

\bibitem{dowker} 
  J.~S.~Dowker and R.~Critchley,
  %``Effective Lagrangian and Energy Momentum Tensor in de Sitter Space,''
  Phys.\ Rev.\ D {\bf 13}, 3224 (1976).

\bibitem{Ebert:2008tp} 
  D.~Ebert, K.~G.~Klimenko, A.~V.~Tyukov and V.~C.~.Zhukovsky,
  %``Pion condensation of quark matter in the static Einstein universe,''
  Eur.\ Phys.\ J.\ C {\bf 58}, 57 (2008)
%  [arXiv:0804.0765 [hep-ph]].

\bibitem{Flachi:2010yz} 
  A.~Flachi and T.~Tanaka,
  %``Chiral Modulations in Curved Space I: Formalism,''
  J. High Energy Phys. 02 (2011) 026;
%  [arXiv:1012.0463 [hep-th]].
%\bibitem{Flachi:2011zr} 
  A.~Flachi,
  %``Chiral Modulations in Curved Space II: Conifold Geometries,''
  J. High Energy Phys. 01 (2012) 023.
%  [arXiv:1111.4131 [hep-th]].

\bibitem{chandra}
 S.~Chandrasekhar, ``\textit{The Mathematical Theory of Black Holes}'', 
 (Oxford University Press, New York, USA, 1983).

\bibitem{lebellac}
 M.~Le~Bellac,
 ``\textit{Thermal Field Theory}'',
 (Cambridge University Press, Cambridge, England, 1996).

\bibitem{Hidaka:2006gd} 
  Y.~Hidaka and M.~Kitazawa,
  %``Chiral transition and mesonic excitations for quarks with thermal masses,''
  Phys.\ Rev.\ D {\bf 75}, 011901 (2007); Y.~Hidaka and M.~Kitazawa, {\it ibid.} 
{\bf 75}, 099901(E) (2007).
%  [hep-ph/0610374].

\bibitem{gorbar}
E.~V.~Gorbar,
  %``Dynamical symmetry breaking in spaces with constant negative curvature,''
  Phys.\ Rev.\ D {\bf 61}, 024013 (1999).
%  [hep-th/9904180].

\bibitem{Yamamoto:2014vda} 
  A.~Yamamoto,
  %``Lattice QCD in curved spacetimes,''
  arXiv:1405.6665 [hep-lat].

\bibitem{Fukushima:2011jc} 
  K.~Fukushima,
  %``QCD matter in extreme environments,''
  J.\ Phys.\ G {\bf 39}, 013101 (2012);
%  [arXiv:1108.2939 [hep-ph]].
%\bibitem{Fukushima:2013rx} 
  K.~Fukushima and C.~Sasaki,
  %``The phase diagram of nuclear and quark matter at high baryon density,''
  Prog.\ Part.\ Nucl.\ Phys.\  {\bf 72}, 99 (2013).
%  [arXiv:1301.6377 [hep-ph]].

\bibitem{Fukushima:2003fw} 
  K.~Fukushima,
  %``Chiral effective model with the Polyakov loop,''
  Phys.\ Lett.\ B {\bf 591}, 277 (2004).
%  [hep-ph/0310121].

\bibitem{Sasagawa:2012mn} 
  S.~Sasagawa and H.~Tanaka,
  %``The separation of the chiral and deconfinement phase transitions in the curved space-time,''
  Prog.\ Theor.\ Phys.\  {\bf 128}, 925 (2012).
%  [arXiv:1209.2782 [hep-ph]].

\bibitem{Flachi:2013zaa} 
  A.~Flachi,
  %``Dual Fermion Condensates in Curved Space,''
  Phys.\ Rev.\ D {\bf 88}, 085011 (2013).
%  [arXiv:1304.6880 [hep-th]].

\bibitem{Elizalde:1996am} 
  E.~Elizalde, S.~D.~Odintsov, and A.~Romeo,
  %``Effective potential for a covariantly constant gauge field in curved space-time,''
  Phys.\ Rev.\ D {\bf 54}, 4152 (1996);
%  [hep-th/9607189].
%\bibitem{Kharzeev:2004ct} 
  D.~Kharzeev, E.~Levin, and K.~Tuchin,
  %``QCD in curved space-time: A Conformal bag model,''
  Phys.\ Rev.\ D {\bf 70}, 054005 (2004).
%  [hep-ph/0403152].

\bibitem{Braun:2007bx} 
  J.~Braun, H.~Gies, and J.~M.~Pawlowski,
  %``Quark Confinement from Color Confinement,''
  Phys.\ Lett.\ B {\bf 684}, 262 (2010).
%  [arXiv:0708.2413 [hep-th]].

\end{thebibliography}
\end{document}